\documentclass{article}

\begin{document}

\centerline{\huge  The Holography of Gravity encoded in a relation}
\centerline{\huge  between Entropy, Horizon area and Action for gravity}

\bigskip

\centerline{\large T. Padmanabhan}
\centerline{\large IUCAA,  Post Bag 4, Ganeshkhind, Pune - 411 007}
\centerline{\large email: nabhan@iucaa.ernet.in}

\bigskip

\begin{abstract}

 I provide a  general proof of the conjecture that one can attribute an entropy to  the area of {\it any}  horizon.
 This is done
by  constructing  a canonical
ensemble of a subclass of  spacetimes with a fixed value for the temperature $T=\beta^{-1}$ and evaluating the {\it exact} partition function $Z(\beta)$. For  spherically symmetric spacetimes with a horizon at $r=a$,  the  partition function has the generic form
$Z\propto \exp[S-\beta E]$, where $S= (1/4) 4\pi a^2$  and $|E|=(a/2)$.  Both $S$ and $E$ are determined entirely by the properties of the metric near the horizon. This analysis reproduces the conventional result for the black-hole spacetimes and provides a simple and consistent interpretation of entropy and energy for De Sitter spacetime. For the Rindler spacetime the entropy per unit transverse area turns out to be $(1/4)$ while the energy is zero.  Further,
 I show  that the relationship between entropy and area allows one to construct the action for the gravitational field on the bulk and 
thus the full theory.  In this sense, gravity is intrinsically holographic. 
\end{abstract}

\noindent

Among the class of Lorentzian spacetime metrics which allow a positive definite continuation to the Euclidean time coordinate $\tau=it$, there exists a subclass of spacetime metrics which require $\tau$ to be treated as periodic with some period $\beta$. It is natural to interpret such a feature as describing
a    finite temperature field theory with temperature $T=\beta^{-1}$. 
(For a review, see e.g., [1].) 
In the case of black hole spacetimes, one can also associate an entropy with the horizon and construct
a consistent parallel with thermodynamics. While there is some indication that we can associate an entropy with
the area of {\it any } horizon (Rindler, De Sitter ......),  any such
association  will also require defining the energy for such a spacetime in order to provide  consistent 
thermodynamic relationships. This, however, is not an easy task in general relativity and hence progress has been somewhat limited in this issue. 

However, if the association of thermodynamical variables with horizons is of truly fundamental significance, then it is indeed necessary that our conclusions are applicable to {\it any} horizon and  it must be possible to attribute an entropy to
{\it any} horizon. In fact, if the entropy of spacetimes arise because some information is hidden by the horizon, then all horizons (even the observer dependent ones
like Rindler or De Sitter horizons) must have an entropy.  Further, one must be able to obtain such a result in an
elegant manner, from standard statistical mechanical procedures. That is, results must `` fall out" of a proper argument
allowing us to: (i) associate entropy with any horizon and (ii) identify the energy content of the spacetime.  I will now
show that this is indeed possible. What is more, one can use the relationship between the entropy and area of horizon as the starting point and reconstruct the full gravitational Lagrangian of the theory. This suggests a deep, holographic, relationship between the surface terms in general relativity  and the theory on the bulk.

A wide class of such spacetimes with horizons, analysed in the literature, has the form
    \begin{equation}
     ds^2=f(r)dt^2-f(r)^{-1}dr^2 -dL_\perp^2
     \label{basemetric}
     \end{equation}
  where $f(r)$ vanishes at some surface $r=a$, say, with $f'(a)\equiv B$ remaining finite. When $dL_\perp^2$
  is taken as the metric on 2-sphere and $r$ is interpreted as the radial coordinate $[0\leq r\leq \infty]$, equation (\ref{basemetric}) covers a variety of spherically symmetric spacetimes  (including Schwarzschild, Reissner-Nordstrom, De Sitter etc.) with a compact horizon at $r=a$. If $r$ is interpreted as one of the Cartesian coordinates $x$ with $(-\infty\leq x\leq \infty)$ and $dL_\perp^2=dy^2+dz^2, f(x)=1+2gx,$ equation (\ref{basemetric}) can describe the Rindler frame in flat spacetime. We shall first concentrate on compact horizons with $r$ interpreted as radial coordinate,
and comment on the Rindler frame at the end.

 Since the metric is static, Euclidean continuation is trivially effected by $t\to
  \tau=it$ and an examination of the conical singularity near $r=a$ [where $f(r) \approx B(r-a)$] shows that $\tau$ should be interpreted as periodic with period $\beta=4\pi/|B|$ corresponding to the temperature $T=|B|/4\pi$. One can prove quite rigorously $^{[1,2]}$ that the spacetime described by (\ref{basemetric}) is endowed with this temperature which
  --- in turn --- depends
  {\it only on the behaviour of the metric near the horizon}. The form of $f(r)$ is arbitrary except for the constraint that $f(r)\approx B (r-a)$ near $r=a$.
  
  The next logical question will be whether one can associate other thermodynamic quantities, especially the entropy, with such spacetimes.$^3$ Given that the temperature can be introduced very naturally, just using the behaviour of metric near the horizon, one would look for a similarly elegant and natural derivation of the entropy. To achieve this, I begin by noting that the
 class of metrics in
  (\ref{basemetric}) with the behaviour $[f(a)=0,f'(a)=B]$ constitute a canonical ensemble at constant temperature  since they all have the same temperature $T=|B|/4\pi$ . The partition function for this ensemble is given  by the path integral sum
   \begin{eqnarray}
    Z(\beta)&=&\sum_{g\epsilon {\cal S}}\exp (-A_E(g))  \\
 &=&\sum_{g\epsilon {\cal S}}\exp \left(-{1\over 16\pi}\int_0^\beta  d\tau \int d^3x \sqrt{g_E}R_E[f(r)]\right)\nonumber
     \label{zdef}
     \end{eqnarray}
  where I have made the Euclidean continuation of the Einstein action and imposed the periodicity in $\tau$
  with period $\beta=4\pi/|B|$.
  The sum is restricted to the set ${\cal S}$ of all metrics of the form in 
  (\ref{basemetric}) with the behaviour $[f(a)=0,f'(a)=B]$ and the Euclidean Lagrangian is a functional of $f(r)$.
  No source term or cosmological constant (which cannot be distinguished from certain form of source) is included since the idea is to obtain a result which depends purely on the geometry.
   The spatial integration will be restricted to a region bounded by the 2-spheres $r=a$ and $r=b$, where
  the choice of $b$ is arbitrary except for the requirement that  within the region of integration the Lorentzian
  metric must have the proper signature with $t$ being a time coordinate. The remarkable feature is the form of the
  Euclidean action for this class of spacetimes.  Using the result
  \begin{equation}
  R=\nabla_r^2 f -{2\over r^2}{d\over dr}\left[r(1-f)\right]\label{reqn}
  \end{equation}
  valid for metrics of the form in (\ref{basemetric}), a
   straight forward calculation shows that
   \begin{eqnarray}
  - A_E&=&{\beta\over 4}\int_a^b dr\left[-[r^2f']'+2[r(1-f)]'\right] \nonumber\\
&=&{\beta\over 4}[a^2B -2a]+Q[f(b),f'(b)]
     \label{zres}
     \end{eqnarray}
where $Q$ depends on the behaviour of the metric near $r=b$ and we have used the 
conditions $[f(a)=0,f'(a)=B]$. The sum in (\ref{zdef}) now reduces to summing over the values of $[f(b),f'(b)]$
with a suitable (but unknown) measure. This sum, however, will only lead to a factor which we can ignore in deciding about the dependence of $Z(\beta)$ on the form of the metric near $r=a$. Using $\beta=4\pi/B$
(and taking $B>0$, for the moment)  the final result can be written in a very suggestive form:
 \begin{equation}
  Z(\beta)=Z_0\exp \left[{1\over 4}(4\pi a^2) -\beta({a\over 2})  \right]\propto 
  \exp \left[S(a) -\beta E(a)  \right]
     \label{zresone}
     \end{equation}
with the identifications for the entropy and energy being given by:
\begin{equation}
S={1\over 4} (4\pi a^2) = {1\over 4} A_{\rm horizon}; \quad E = {1\over 2} a = \left( {A_{\rm horizon}\over 16 \pi}\right)^{1/2}
\end{equation}
  In addition to the simplicity, 
     the following features are noteworthy regarding this result:
     
    (i) The result is local in the sense that it depends only on the form of the metric near the 
    horizon.  In particular, the definition of energy does not depend on the asymptotic flatness of the metric.
    
    (ii) The partition function was evaluated with two very natural conditions: $f(a) =0$ making 
    the surface $r=a$ a compact horizon and $f'(a) = $ constant which is the 
    proper characterization of the canonical ensemble of spacetime metrics. Since temperature
    is well defined for the class of metrics which I have considered, this canonical ensemble is defined without any   ambiguity.  This allows me to  sum over a  class of spherically symmetric spacetimes at one go rather than
    deal with, say, black-hole spacetimes and De Sitter spacetime separately. Unlike many of the 
previous approaches,   I do {\it not} evaluate the
path integral in the WKB limit, confining to metrics which are solutions of Einstein's equations. 
 (When the path integral sum is evaluated  using WKB ansatz for vacuum spacetimes like Schwarzschild black-hole 
--- as, e.g., in the  works by Gibbons and Hawking$^4$ --- the scalar curvature
$R$ vanishes and a surface contribution arises from the trace
 of the second fundamental form on the boundary. 
The surface contribution which arises in (\ref{zres}) is different.) Conceptually,
a canonical ensemble for a minisuperspace of metrics  of the form in (\ref{basemetric})  should 
be constructed by keeping the temperature constant {\it without} assuming the metrics
to be the solutions of Einstein's equation; this is what I do and exploit the form of $R$ given by (\ref{reqn}).
Since this action involves second derivatives, it is not only allowed but even required to fix both $f$ and $f'$
at the boundaries.

    (iii) In the case of the Schwarzschild black-hole with $a=2M$, the energy turns out to
    be $E=(a/2) = M$ which is as expected. (More generally,
	$E=(A_{\rm horizon}/16\pi)^{1/2}$ corresponds to the so called `irreducible mass' in 
       BH spacetimes$^5$). Of course, the identifications $S=(4\pi M^2)$,
    $E=M$, $T=(1/8\pi M)$ are consistent with the result $dE = TdS$ in this particular case.
    
    Most importantly, our analysis provides an interpretation of entropy and energy in the case
    of De Sitter universe which is gaining in popularity. In this case, $f(r) = (1-H^2r^2)$, $a=H^{-1}, B=-2H$.
Since the region where $t$ is time-like is ``inside'' the horizon, the integral for $A_E$ in (\ref{zres}) should be taken from some arbitrary value $r=b$ to $r=a$ with $a>b$. So the horizon contributes in the upper limit of the integral
introducing a change of sign in (\ref{zres}). Further, since $B<0$, there is another negative sign in the area term
from $\beta B\propto B/|B|$. Taking all these into account we get, in this case, 
\begin{equation}
  Z(\beta)=Z_0\exp \left[{1\over 4}(4\pi a^2) +\beta({a\over 2})  \right]\propto 
  \exp \left[S(a) -\beta E(a)  \right]
     \label{zrestwo}
     \end{equation}
giving
$S=(1/ 4) (4\pi a^2) = (1/ 4) A_{\rm horizon}$ and  $E=-(1/2)H^{-1}$. 
These definitions do satisfy the relation $TdS -PdV =dE$ when it is noted that the De Sitter universe has 
    a non zero pressure $P=-\rho_\Lambda=-E/V$ associated with the cosmological constant. In fact,
if we use the ``reasonable" assumptions $S=(1/4)(4\pi H^{-2}), V\propto H^{-3}$ and $E=-PV$ in the equation
$TdS -PdV =dE$ and treat $E$ as an unknown function of $H$, we get the equation $H^2(dE/dH)=-(3EH+1)$
which integrates to give precisely $E=-(1/2)H^{-1}. $ 

 Let us now consider the spacetimes with planar symmetry for which (\ref{basemetric}) is still applicable with $r=x$ being a Cartesian coordinate and $dL_\perp^2=dy^2+dz^2$. In this case $R=f''(x)$ and the action becomes
\begin{eqnarray}
-A_E&=&-{1\over 16\pi}\int_0^\beta d\tau\int dy dz \int_a^b dx f''(x)\nonumber\\
&=& {\beta\over 16\pi} A_\perp f'(a)+Q[f'(b)]
\label{rindleraction}
\end{eqnarray}
where we have confined the transverse integrations to a surface of area $A_\perp$. If we now sum over
all the metrics with $f(a)=0,f'(a)=B$ and $f'(b)$ arbitrary, the partition function will become
\begin{equation}
Z(\beta)=Z_0\exp({1\over 4}A_\perp)
\end{equation}
which suggests that planar horizons have an entropy of (1/4) per unit transverse area but zero energy. This includes
Rindler frame as a special case. Note that if we freeze  $f$ to its Rindler form $f=1+2gx$, 
(by demanding the validity of Einstein's equations in the WKB approach, say)
then $R=f''=0$ as it should.
In the action in (\ref{rindleraction}), $f'(a)-f'(b)$ will give zero. It is only because I am
 {\it not} doing a WKB analysis --- but 
 varying $f'(b)$ with fixed $f'(a)$ --- that I obtain an entropy for these spacetimes.  
 
 It is straight forward to use this approach in $D$ dimensions with the hope that insights gained in $D\neq 4$ may be of some help. In $D=(1+2)$, for example, metrics of the type in (\ref{basemetric}) with $dL^2_\perp=r^2 d\theta^2$ will give $S=(1/4)(2\pi a)=(1/4)A_{\rm horizon}$ with $E=0$. The vanishing of energy  signifies the fact that at the level of the metric, Einstein's equations are vacuous in (1+2) and we have not incorporated any topological
effects [like deficit angles corresponding to point masses in (1+2) dimensions] in our approach.

Finally, let me indicate a deeper connection between this result and the holographic nature of gravity.  
To do this I raise the status of the above results to that of a postulate: {\it The dynamics of the gravity must be described by
an action such that, in static spacetimes with horizons which leads to periodicity in imaginary time, the action has a surface contribution which is one quarter of the area
of the horizon.} This requires the surface contribution of the action on a horizon ${\cal H}$ to be of the form
\begin{equation}
A=\int_{\cal H} {d^4x \over 16\pi}L_{grav}\equiv{\beta\over 16\pi}\int_{\cal H} d^3x[\partial_aP^a + \cdots]={\beta\over 16\pi}\int_{\cal H} d^3x[\partial_a(\sqrt{-g}\partial_b g^{ab}) + \cdots]
\end{equation}
where  $\beta$ is the period in the imaginary time
arising due to the existence of the horizon and the dots represent the terms which vanish on the horizon. These extra
terms can, however,  be determined uniquely from the requirement of general covariance on the surface; we get
\begin{equation}
P^a=\sqrt{-g}\partial_bg^{ab}+2g^{ab}\partial_b\sqrt{-g}
\label{defofp}
\end{equation}
It is easy to verify that for the spacetimes of the
form in (\ref{basemetric}), the first term in (\ref{defofp}) for $P^r$ will correctly reproduce the entropy as a quarter of the horizon area while
the second term vanishes on the horizon.
The structure of $P^a$ immediately suggests that $L_{grav}$ will contain second derivatives of the metric.
Given any Lagrangian $L(\partial q, q)$ involving only up to the first derivatives of the dynamical variables, it is 
{\it always} possible
to construct another Lagrangian $L'(\partial^2q,\partial q,q)$, involving second derivatives such that it describes the same dynamics. The prescription is:
\begin{equation}
L'=L-{d\over dt}\left(q{\partial L\over \partial\dot q}\right)
\label{lbtp}
\end{equation}
While varying the $L'$, one keeps the {\it momenta} $(\partial L/\partial\dot q)$ fixed
at the endpoints rather than $q'$s. [This result has a simple interpretation in terms of the path integral prescription
in quantum theory, in which the extra term arises while Fourier transforming from $q$ to $p$; see ref.6, page 171.].
What is more, by equating the surface terms to $(\partial L/\partial\dot q)q$ one can obtain $L$. In the case of
gravity, if the (unknown) first order Lagrangian is $L_1(\partial g,g)$  then the  field momenta are
$\pi^{abc} = \left(\partial{\cal L}_1/\partial g_{ab,c}\right)$ where $g_{ab,c} =
 \partial
g_{ab}/\partial x^c$. Since the term which is fixed at the surface is given by 
$(g_{ab}\partial{\cal L}_1/\partial g_{ab,c})$ we can integrate the equation
 \begin{equation}
 \left({\partial{\cal L}_1\over\partial g_{ab,c}}g_{ab}\right)=P^c=\sqrt{-g}\partial_bg^{cb}+2g^{cb}\partial_b\sqrt{-g}
\end{equation}
to obtain
the first order Lagrangian density (see ref. 6: page 326):
\begin{equation}
{\cal L}_1 \equiv \sqrt{-g}{\cal G}= \sqrt{-g} \, g^{ik} \left(\Gamma^m_{i\ell}\Gamma^\ell_{km} -
\Gamma^\ell_{ik} \Gamma^m_{\ell m}\right).
\end{equation}  
Following the prescription of  (\ref{lbtp}) we now
subtract $\partial (g_{ab} \pi^{abc})/\partial x^c$ from ${\cal L}_1$ to get
the equivalent Lagrangian $L_{grav}$ with second derivatives, which {\it turns out} to be the standard 
Einstein-Hilbert lagrangian:
  \begin{equation}
L_{grav}= {\cal L}_1 - {\partial\over\partial x^c} \left(g_{ab}
\pi^{abc}\right) = R\sqrt{-g}.
\end{equation}
Thus the surface terms dictate the form of the Einstein Lagrangian in the bulk.

The above analysis shows that the postulate of gravitational action being equal to one quarter of the area of the horizon,
added to the requirement of general covariance, uniquely determines the gravitational action principle.  In other words,
the idea that surface areas of horizons encode one quarter bit of information per Planck area allows one to determine the nature of gravitational interaction on the bulk, which is an interesting realization of the holographic principle.

\end{document}